\documentclass[superscriptaddress,nofootinbib,showkeys,showpacs,byrevtex]{revtex4}
\usepackage{graphicx}
\usepackage{epstopdf}

\begin{document}      
\title{Production of the Pentaquark Exotic Baryon $\Xi_5$ in
$\bar{K}N$ Scattering:\\ $\bar{K}N\to K\Xi_5$
and $\bar{K}N\to K^{*}\Xi_5$}
\author{Seung-il Nam}
\email[E-mail: ]{sinam@yukawa.kyoto-u.ac.jp}
\affiliation{Yukawa Institute for Theoretical Physics (YITP), Kyoto
University, Kyoto 606-8502, Japan}  
\author{Atsushi Hosaka}
\email[E-mail: ]{hosaka@rcnp.osaka-u.ac.jp}
\affiliation{Research Center for Nuclear Physics (RCNP), Ibaraki, Osaka
567-0047, Japan}
\author{Hyun-Chul Kim}
\email[E-mail: ]{hchkim@pusan.ac.kr}
\affiliation{Department of Physics, Inha University, Incheon 402-751,
Republic of Korea.} 
\affiliation{Department of
Physics and Nuclear Physics \& Radiation Technology Institute (NuRI),
Pusan National University, Busan 609-735, Korea} 
\date{\today}
\begin{abstract}
We investigate the production of the 
newly found pentaquark exotic baryon $\Xi_5$ in
the $\bar{K}N\to K\Xi_5$ and
the $\bar{K}N\to K^{*}\Xi_5$ reactions at the tree level. 
We consider both positive- and negative-parities of the $\Xi_5$.  
The reactions are dominated by the $s$- and the $u$-channel processes, 
and the resulting cross sections are observed to depend 
very much on the parity of 
$\Xi_5$ and on the type of form factor.  
We have seen that the cross sections 
for the positive-parity $\Xi_5$ are generally about 
a hundred times 
larger than those of the negative-parity one.  
This large difference in the cross sections
will be useful for further study of the pentaquark baryons.  
\end{abstract}
\maketitle
\section{introduction}
The experimental observation of the $\Theta^+$ performed by the LEPS 
collaboration at SPring-8~\cite{Nakano:2003qx}, which is motivated by
Diakonov {\em et al.}~\cite{Diakonov:1997mm}, has paved the way for 
intensive studies on the exotic five-quark baryon 
states, also known as {\it pentaquarks}, 
experimentally~\cite{experiment} as well as 
theoretically~\cite{Praszalowicz:2003tc,Jaffe:2003sg,
Hosaka:2003jv,Kondo:2004rn,Stancu:2003if,Glozman:2003sy,Huang:2003we,Sugiyama:2003zk,
Zhu:2003ba,Sasaki:2003gi,Csikor:2003ng,Liu:2003rh,Hyodo:2003th,
Oh:2003kw,nam1,nam2,nam3,nam4,Thomas:2003ak,Hanhart:2003xp,
Liu:2003zi,Zhao:2003gs,Yu:2003eq,Kim:2003ay,Huang:2003bu,Liu:2003ab,
Li:2003cb}.  As a consequence of the finding of the $\Theta^{+}$, the
existence of other pentaquark baryons, such as the $N_5$,
$\Sigma_5$, and $\Xi_5$, which have also been predicted theoretically, is anticipated.  

The NA49~\cite{Alt:2003vb} collaboration reported a signal for the pentaquark baryon $\Xi_5$, which was also predicted theoretically. The $\Xi_5$ was found to have a mass of
$1862\,{\rm MeV}$, a strangeness $S = -2$, and an isospin 
$I = 3/2$.  It is characterized by its narrow decay width of $\sim
18\,{\rm MeV}$, like that of the $\Theta^+$.  However, we have thus far no
concrete experimental evidence for its quantum numbers such as {\it
spin} and {\it parity}.   

As for the parity of the $\Theta^+$, a consensus has not been reached.  For example, the chiral soliton
model~\cite{Diakonov:1997mm,Praszalowicz:2003tc}, 
the diquark model~\cite{Jaffe:2003sg}, the chiral potential
model~\cite{Hosaka:2003jv}, and constituent quark models with
spin-flavor
interactions~\cite{Stancu:2003if,Glozman:2003sy,Huang:2003we} prefer a positive-parity for the $\Theta^+$ whereas the QCD 
sum rule approach~\cite{Sugiyama:2003zk,Zhu:2003ba}~\footnote{
Recently, Ref.~\cite{Kondo:2004rn} pointed out that the exclusion of the 
non-interacting $KN$ state from the two-point correlation function may 
reverse the parity of $\Theta^+$.}
and the quenched 
lattice
QCD~\cite{Sasaki:2003gi,Csikor:2003ng} have supported a negative-parity.  
In the meanwhile, various reactions for 
$\Theta^{+}$ production~\cite{Liu:2003rh,Hyodo:2003th,Oh:2003kw,
nam1,nam2,nam3,nam4,Thomas:2003ak,Hanhart:2003xp,Liu:2003zi,
Zhao:2003gs,Yu:2003eq,Kim:2003ay,Huang:2003bu,Liu:2003ab,Li:2003cb} 
have been investigated, where the determination of the parity 
of $\Theta^+$ has been  emphasized.  
In many cases, the total cross-sections of the positive-parity $\Theta^+$
production is typically about ten times larger than those of
the negative-parity one. Liu {\it et al.}~\cite{Liu:2003dq} evaluated the $\gamma N\to K\Xi_5$ reactions, assuming the positive-parity
$\Xi_5$ and its spin $J=1/2$.  However, since the parity of the $\Xi_5$ is not known yet, it is worthwhile studying the dynamics of $\Xi_5$ production with two different parities taken into account.   

However, we note that negative results for the pentaquark baryons have emerged recently. Especially, the CLAS collaboration at Jefferson laboratory could not see any obvious evidence for the $\Theta^+$ pentaquark, which was expected to have a peak at about $1530\sim1540$ MeV in the reaction 
$\gamma p \to \bar K^0 \Theta^+$~\cite{DeVita:2005CLAS}. Moreover, the existence of an $S=-2$ penataquark, such as the $\Xi_5$ or the charmed pentaquark ($\Theta^+_c$), has not been completely confirmed yet.

In the present work, nonetheless, for the unclear status of the pentaquark,  we want to investigate the $\Xi_5$ production from the 
$\bar{K}N\to K\Xi_5$ and the $\bar{K}N\to
K^{*}\Xi_5$ reactions. Due to the exotic strangeness quantum number the $\Xi_5$ has, the reaction process at the tree level becomes considerably simplified. This is a very specific feature of the process containing the exotic strangeness quantum number. We will follow the same framework as in 
Refs.~\cite{Oh:2003kw,nam1,nam2,nam3,nam4,Zhao:2003gs,Yu:2003eq}.  
We assume that the spin of the $\Xi_5$ is $1/2$~\cite{Liu:2003dq}.  
Then, we estimate the total and the differential cross-sections 
for the production of $\Xi_5$ with positive and negative parities.  

This paper is organized as follows: In Section II, we define the
effective Lagrangians and construct the invariant amplitudes.  In 
Section III, we present the numerical results for the total and the differential
cross-sections for both positive- and negative-parity $\Xi_5$.  
Finally, in Section IV, we briefly summarize our discussions.    

 \section{Effective Lagrangians and amplitudes}
\begin{figure}[t]
\includegraphics[width=10.0cm]{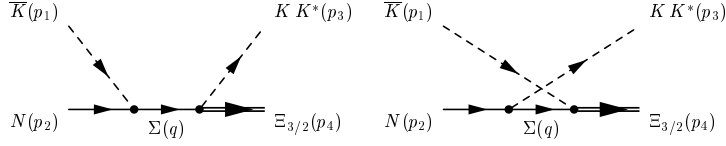}
\caption{Born diagrams, $s$-- (left) and $u$--channels
(right) for $\Xi_5$ productions}
\label{nmset00}
\end{figure}

We study the reactions $\bar{K}N\to K\Xi_5$ 
and $\bar{K}N\to K^{*}\Xi_5$ by using an effective Lagrangian at the tree level of Born diagrams.  
The reactions are schematically presented in 
Fig.~\ref{nmset00}, where we define the four momenta of each 
particle for the reactions by
$p_{1, \cdots 4}$.  
There is no $t$-channel contribution because strangeness-two ($S=2$) mesons do not
exist.  As discussed in Ref.~\cite{Oh:2003fs}, we do not include the 
${B}_{\bar{10}}M_{8}B_{10}$ coupling because it is forbidden in
exact SU(3) flavor symmetry.  
Hence, the interaction Lagrangians can be written as   
\begin{eqnarray}
\mathcal{L}_{KN\Sigma}&=&ig_{KN\Sigma}\bar{\Sigma}
\gamma_{5}KN\,+\,{\rm (h.c.)},
\nonumber \\
\mathcal{L}_{K\Sigma\Xi_5}&=&ig_{K\Sigma\Xi_5}\bar{\Xi}_5
\Gamma_{5}K\Sigma\,+\,{\rm (h.c.)},
\nonumber \\
\mathcal{L}_{K^{*}N\Sigma}&=&g_{K^{*} N\Sigma}\bar{\Sigma}
\gamma_{\mu}K^{*\mu}N\,+\,{\rm (h.c.)}, 
\nonumber \\
\mathcal{L}_{K^{*}\Sigma\Xi_5}&=&g_{K^{*}\Sigma\Xi_5}\bar{\Xi}_5
\gamma_{\mu}\hat{\Gamma}_{5}K^{*\mu}\Sigma\,+\,{\rm (h.c.)}, 
\label{lagrangians} 
\end{eqnarray}
where $\Sigma$, $\Xi_5$, $N$, $K$, and $K^{*}$ denote the corresponding
fields for the octet $\Sigma$, the antidecuplet $\Xi_5$, the nucleon,
the pseudo-scalar $K$, and the vector $K^{*}$, respectively.  The
isospin operators are dropped because we treat the isospin states of
the fields explicitly.  We define $\Gamma_{5} = 
\gamma_{5}$ for the positive-parity $\Xi_5$ whereas 
$\Gamma_{5} = {\bf 1}_{4\times4}$ for the negative-parity 
one.  $\hat{\Gamma}_{5}$ is also defined by   
$\Gamma_{5}\gamma_{5}$ for the vector meson $K^{*}$.  
The values of the coupling constants $g_{KN\Sigma}$ and $g_{K^* N
  \Sigma}$ are taken from the new Nijmegen potential~\cite{stokes} as $g_{KN\Sigma} = 3.54$ and $g_{K^{*}N\Sigma} = -2.99$ 
whereas we assume ${\rm SU(3)}$ flavor symmetry for $g_{K \Sigma
  \Xi_5}$ so that we obtain the relation  
$g_{K \Sigma \Xi_5} = g_{KN\Theta}$~\cite{Oh:2003fs}.  Employing the
  decay width $\Gamma_{\Theta \to KN} = 15$ MeV and  
$M_\Theta = 1540$ MeV, we obtain  
$g_{KN\Theta} = g_{K \Sigma \Xi_5} = 3.77\, (0.53)$ 
for the positive (negative) parity.  
The remaining one, $g_{K^*\Sigma\Xi_5}$, is not known, which we will
discuss in the next section.   

The invariant scattering amplitude for
$\bar{K}N\to K\Xi_5$ can be written as 
\begin{equation}
i\mathcal{M}_{x,K} = 
ig_{K\Sigma \Xi_5}g_{KN\Sigma}F^{2}_{x}(q^2)
\bar{u}(p_4)\Gamma_{5}\frac{
{q}_{x}+M_{\Sigma}}
{q^{2}_{x}-M^{2}_{\Sigma}}\gamma_{5}u(p_2)\, , 
\label{amplitudes1}  
\end{equation}
where $x$ labels either the $s$-channel or the $u$-channel, and the 
corresponding momenta are 
$q_{s}=p_{1}+p_{2}$ and $q_{u}=p_{2}-p_{3}$.  
For $\bar{K}N\to K^{*}\Xi_5$, we have
\begin{eqnarray}
i\mathcal{M}_{s,K^{*}}&=&
g_{K^{*}\Sigma \Xi_5}g_{KN \Sigma}
F^{2}_{s}(q^2)\bar{u}(p_4)\rlap{/}{\epsilon}\hat{\Gamma}_{5}
\frac{\rlap{/}{q}_{s}+M_{\Sigma}}{q^{2}_{s}-M^{2}_{\Sigma}}\gamma_{5}u(p_2),\nonumber\\
i\mathcal{M}_{u,K^{*}}&=&
g_{K^{*}N\Sigma}g_{K\Sigma\Xi_5}F^{2}_{u}(q^2)\bar{u}(p_4){\Gamma}_{5} 
\frac{\rlap{/}{q}_{u}+M_{\Sigma}}{q^{2}_{u}-M^{2}_{\Sigma}}\rlap{/}{\epsilon}u(p_2).
\label{amplitudes2} 
\end{eqnarray}
As indicated in Eq.~(\ref{amplitudes1}), 
the coupling constants are commonly factored out for the 
$s$- and the $u$-channels in the $K$ production.  Therefore, there is no
ambiguity due to the sign of the coupling constants.  On the contrary,
there is such an ambiguity due to the unknown sign of 
$g_{K^*\Sigma\Xi_5}$ in the case of $K^*$ production,  

Since the baryon has an extended structure, we need to introduce a form factor.  We employ the form factor~\cite{nam3}   
\begin{equation}
F_{1}(x) =
\frac{\Lambda^{2}_{1}}{\sqrt{\Lambda^{4}_{1}+(x-M^{2}_{\Sigma})^{2}}}
\label{ff1}
\end{equation} 
in such a way that the singularities appearing in the pole diagrams can
be avoided.  Here, $\Lambda_{1}$ and $M_{\Sigma}$ stand for the cutoff
parameter and the $\Sigma$ mass, respectively.  We set the 
cutoff parameter $\Lambda_{1} = 0.85\,{\rm GeV}$ as in Ref.~\cite{nam3}.  This
value was used to reproduce the cross sections of $\gamma p
\to K^{+}\Lambda$.  In order to verify the dependence of the
form factor, we consider also the three-dimensional form factor 
\begin{eqnarray}
F_{2}(\vec{q}^{2}) = \frac{\Lambda^{2}_{2}}{\Lambda^{2}_{2}+|\vec{q}^{2}|},
\label{ff2}
\end{eqnarray}  
where $\vec{q}$ denotes the three momentum of the external meson.  As for
the cutoff parameter, we set $\Lambda_{2} = 0.5\,{\rm GeV}$, which was deduced
from the $\pi N \to K\Lambda$ reaction~\cite{Liu:2003rh}. 
\begin{figure}[t]
\begin{tabular}{cc}
\includegraphics[width=7.5cm]{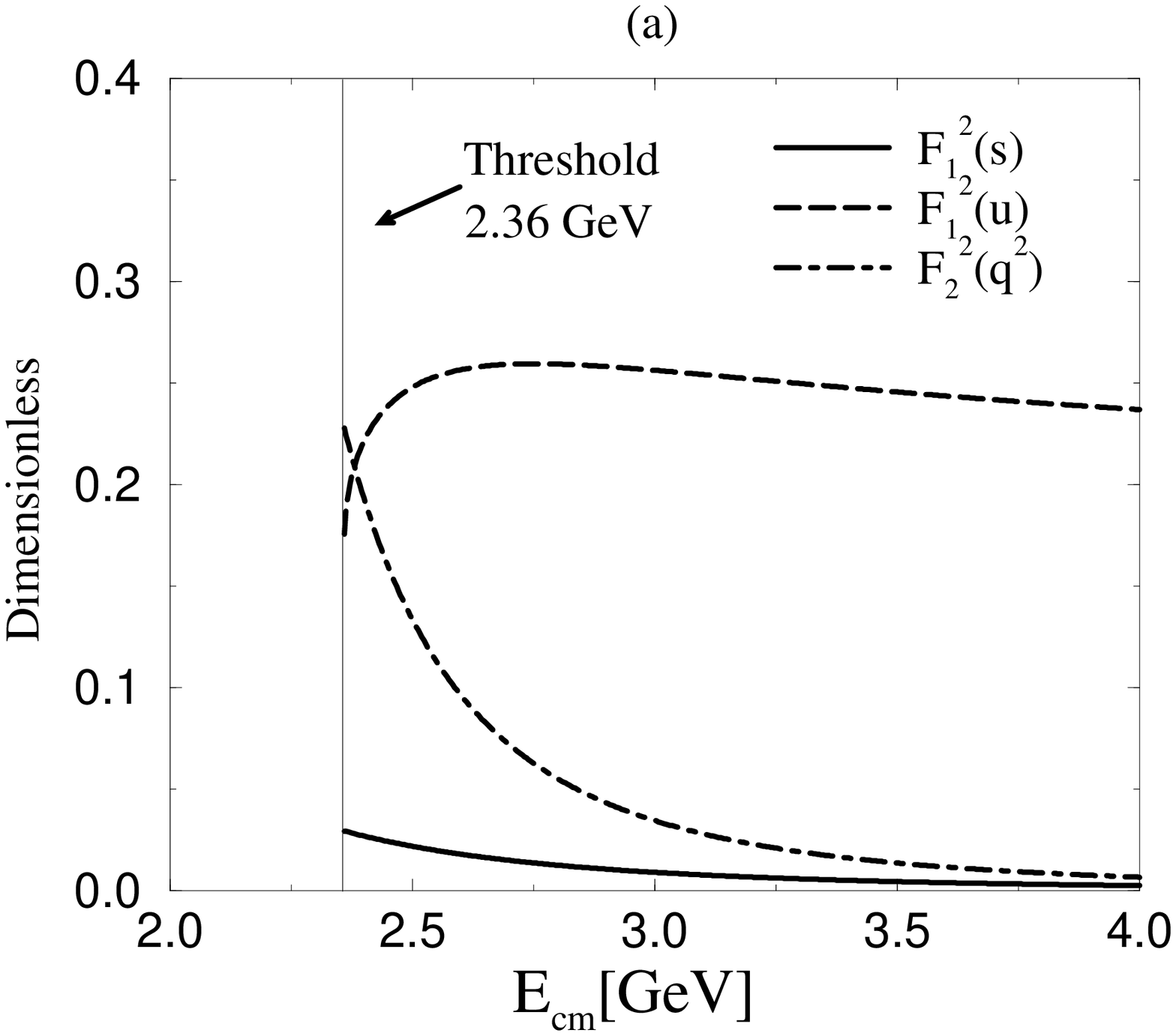}
\includegraphics[width=7.5cm]{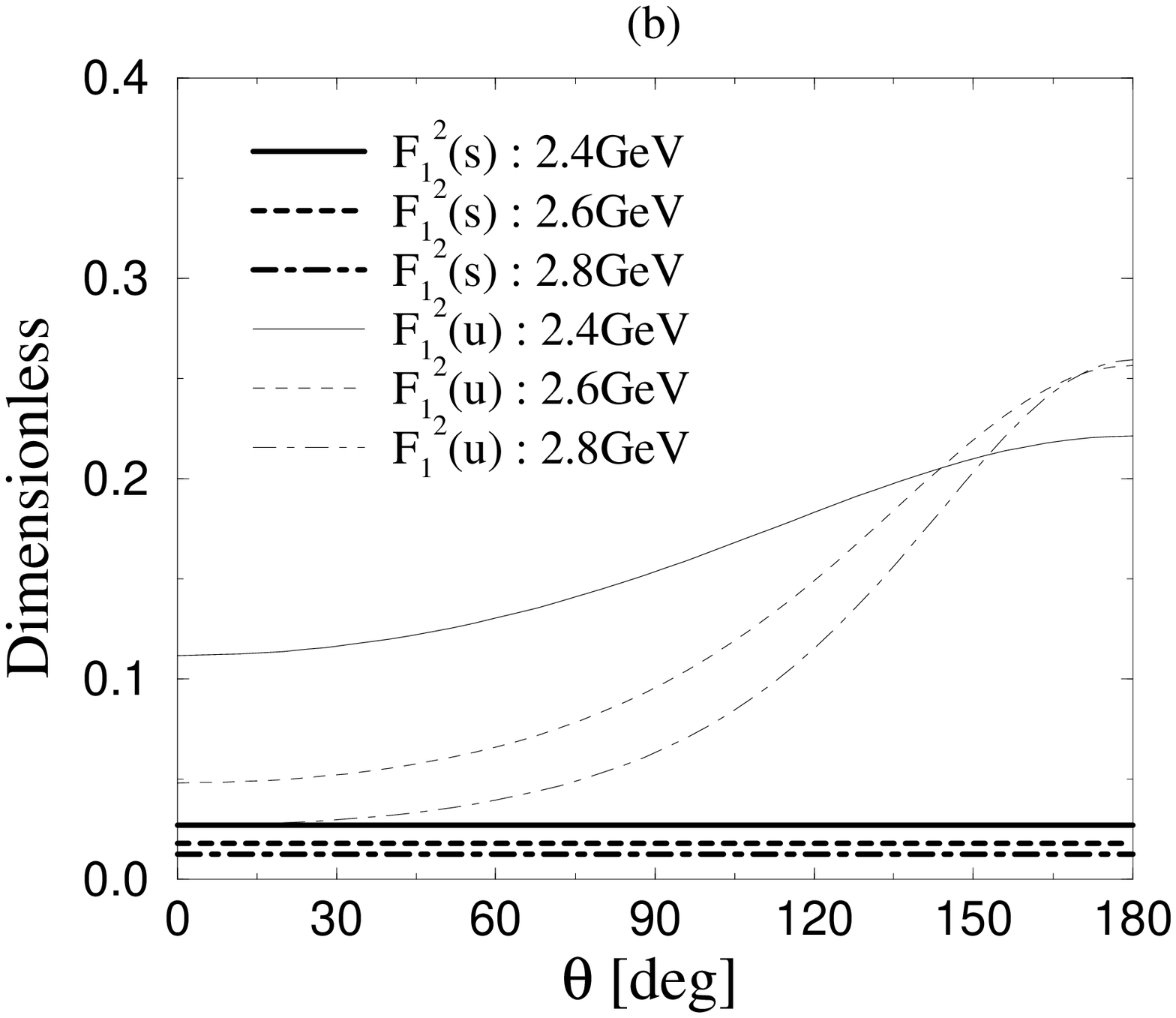}
\end{tabular}
\caption{Energy dependence of the squared form factors 
$F^{2}_1(s)$, $F^{2}_1(u)$ and the $F^{2}_2(\vec q^2)$ (a), and 
angular dependence of the squared form
factor $F^{2}_{1}$ for the $s$- and the $u$-channels at three CM energies. The types of curves are explained by the 
labels in the figure.}      
\label{nmset0} 
\end{figure}

In the left panel (a) of Fig.~\ref{nmset0}, 
we show the dependence of the two form factors 
$F_1$ and $F_2$ on the CM energy while in the left panel (b) the
angular dependence of the $F_1$ form factor is drawn.  The $F_2$ form
factor does not have any angular dependence.  Obviously, they show very
different behaviors.   For instance, the $F_2$ decreases much faster
than $F_1$ as the  center-of-mass (CM) energy grows.  The form
factor $F_{1}$ in the $u$-channel shows a strong enhancement in a
backward direction as the CM energy increases.  As we will see, this
feature has a great effect on the angular dependence of the differential
cross-sections.   
   
\section{Numerical results}
\subsection{$\bar{K}N\to K\Xi_5$}
\begin{figure}[t]
\begin{tabular}{cc}
\includegraphics[width=7.5cm]{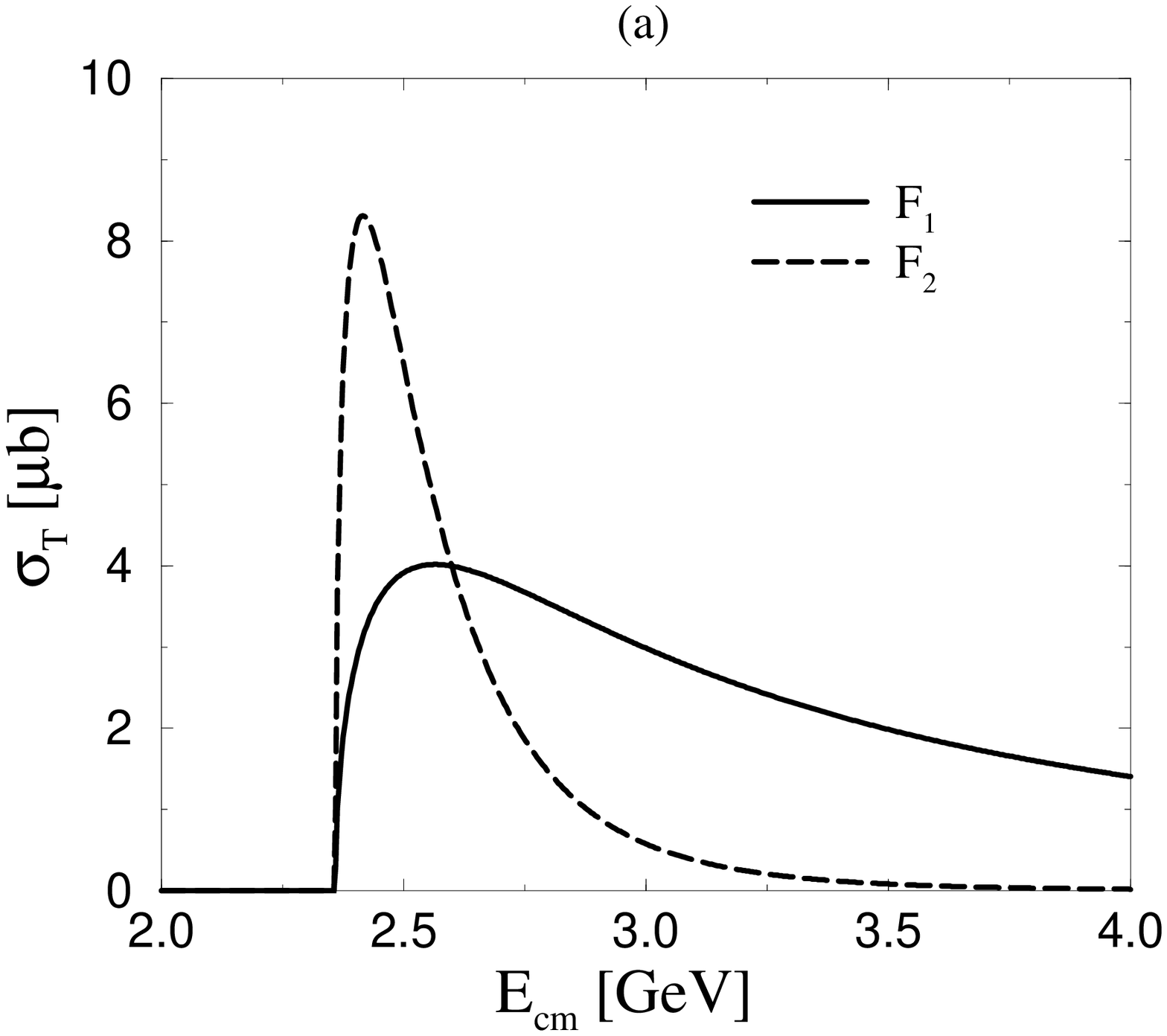}
\includegraphics[width=7.5cm]{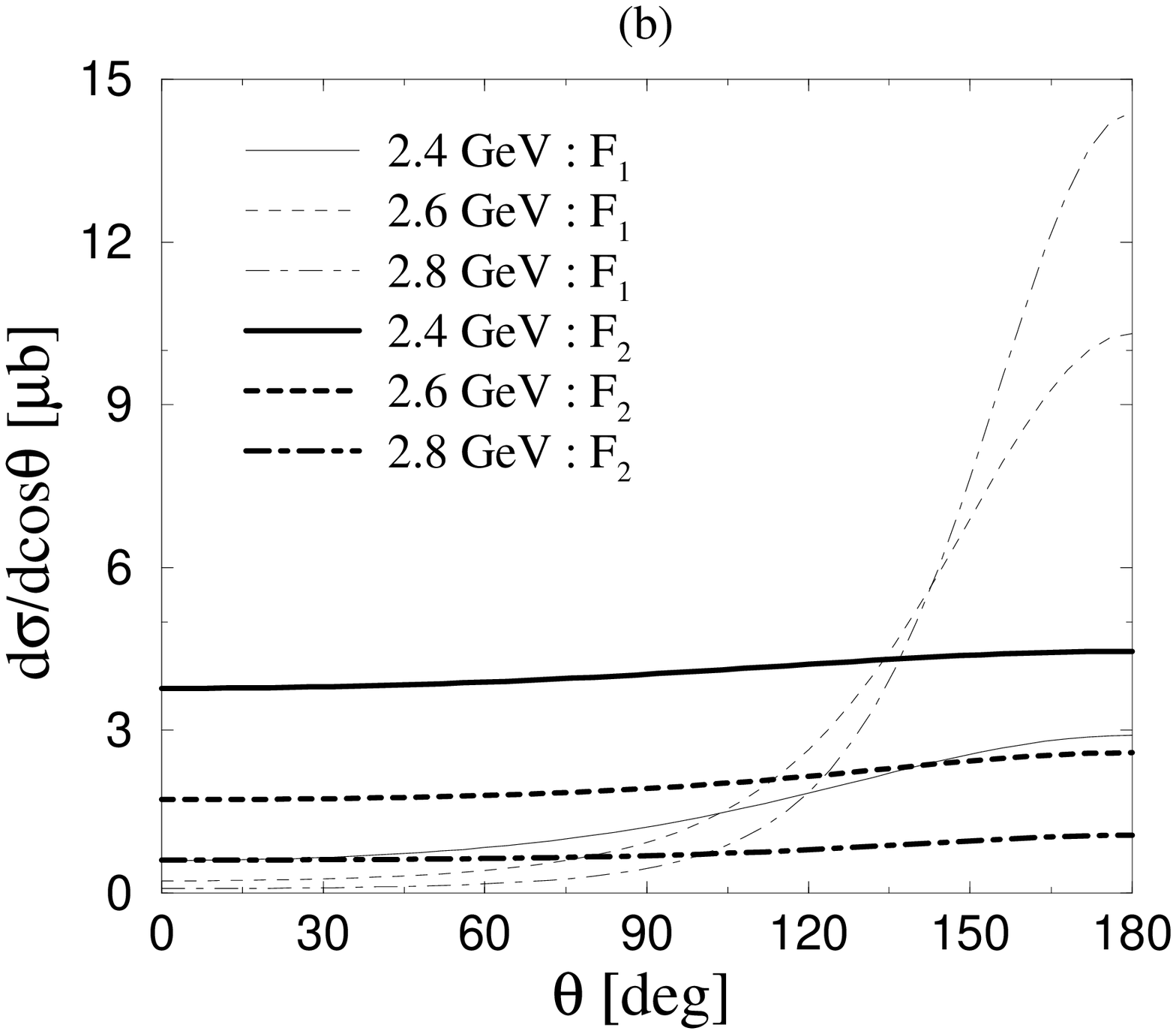}
\end{tabular}
\caption{
Cross sections for production of the positive parity
$\Xi_5$ in the reaction $\bar{K}^{0}p\to K^{0}\Xi^{+}_5$.  
(a) The left panel shows the total cross-sections as functions of 
the center-of-mass energy $E_{\rm CM}$. (b) The right panel shows the differential cross-sections as functions of the 
scattering angle $\theta$ for incident energies 
$E_{\rm CM} = 2.4$, $2.6$, and $2.8$ GeV.  
In both cases, results using the form factors $F_1$ and $F_2$ are shown
as indicated by the labels in the figures. }
\label{nmset1}
\end{figure}

In this subsection, we discuss the results for the reaction 
$\bar{K}N\to K\Xi_5$.  Due to isospin symmetry, we can verify that
the two possible reactions $\bar{K}^{0}p\to
K^{0}\Xi^{+}_5$ and ${K}^{-}n\to K^{+}\Xi^{--}_5$
are exactly the same in the isospin limit.     
In Fig.~\ref{nmset1}, we present the total and the differential   
cross-sections in the left and the right panels, respectively.  The
average values of the total cross-sections are  
$\sigma\sim 2.6\,\mu b$ with the $F_{1}$ form factor and $\sigma\sim 1.5\,\mu b$ with the $F_{2}$ in the energy range  
$E_{CM}^{\rm th} = 2.35 \, {\rm GeV} 
\le
E_{CM}\le 3.35\,{\rm GeV}$ (from the threshold to the point of 
1 GeV larger).  Though the average total cross-sections for the
different form factors are similar in order of magnitude, 
the energy and the angular dependences are very different from 
each other.  They are largely dictated by the form factor, as shown in  
Fig.~\ref{nmset0}.  The angular distributions are drawn in the right
panel (b) of Fig.~\ref{nmset1}, where $\theta$ represents the 
scattering angle between the incident and the final kaons in the CM
system.  We show the results at $E_{CM} = 2.4$, $2.6$, and $2.8$ GeV.   As shown there, when $F_1$ is used, the
backward production is strongly enhanced while the cross sections are
almost flat apart from a tiny increase in the backward region, when $F_2$ is employed.  Note that the angular dependence of the latter
is the same as that of the bare cross section without the form factor.  
\begin{figure}[t]
\begin{tabular}{cc}
\includegraphics[width=7.5cm]{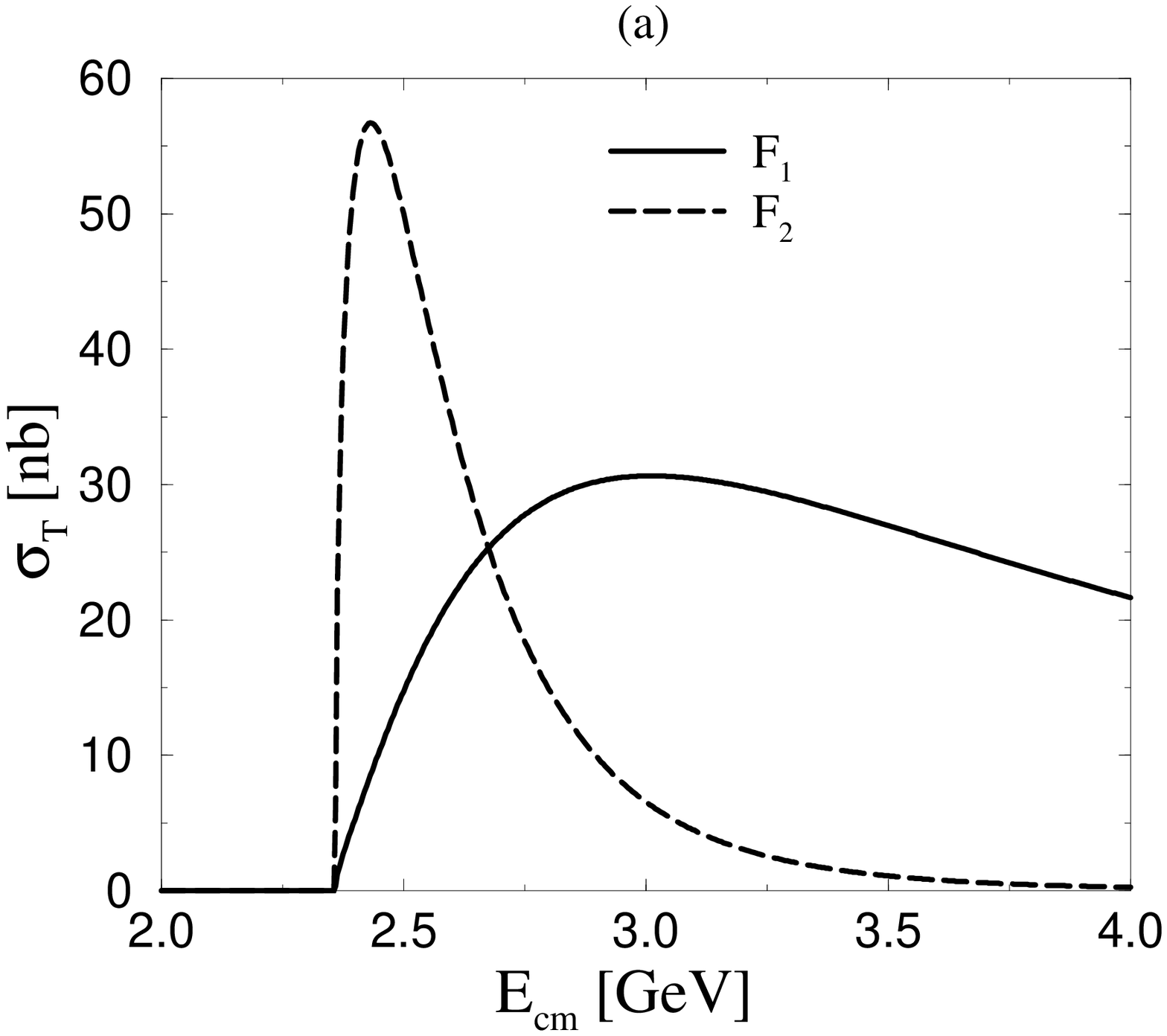}
\includegraphics[width=7.5cm]{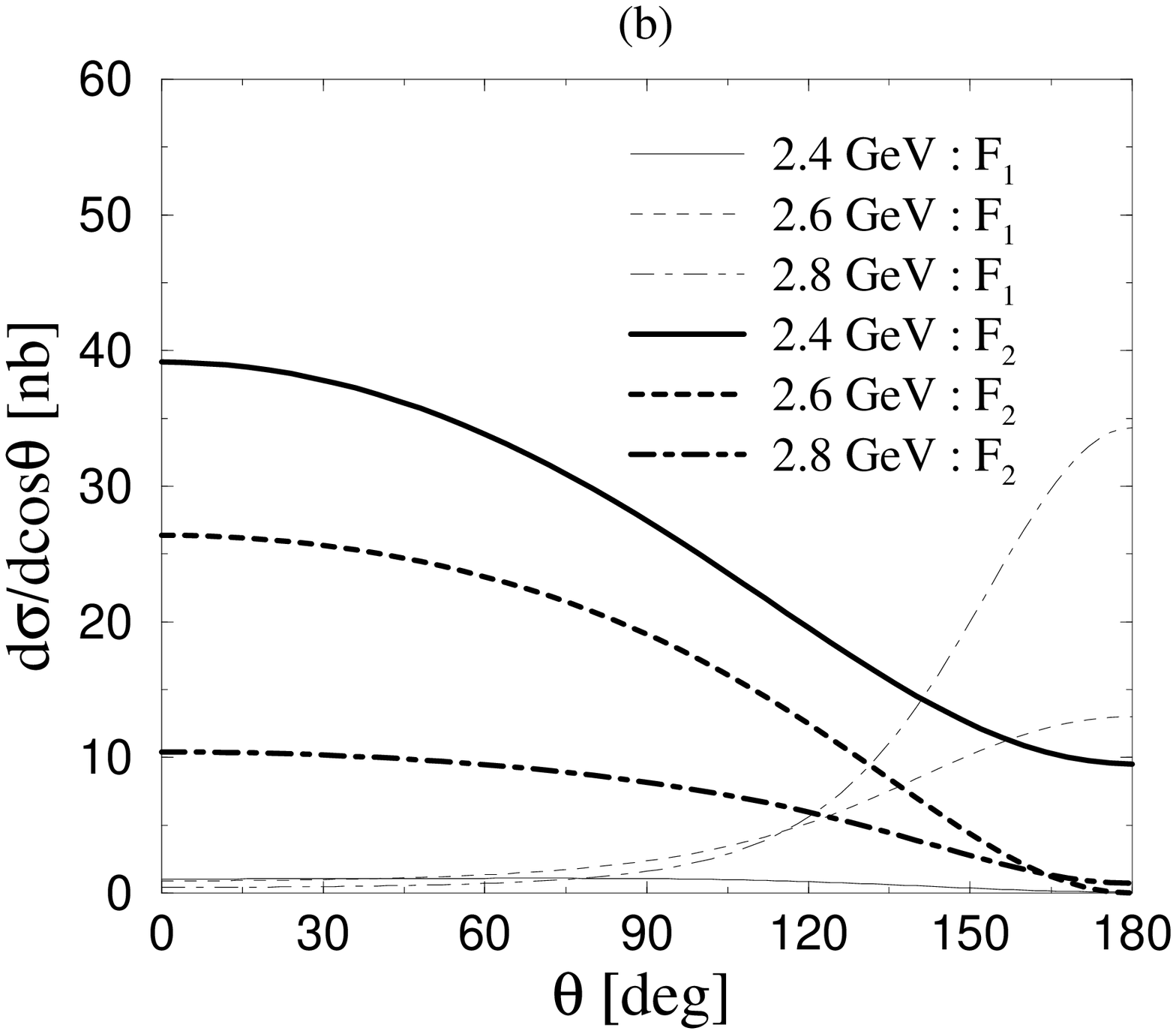}
\end{tabular}
\caption{
Cross sections for production of the negative parity
$\Xi_5$ in the reaction $\bar{K}^{0}p\to K^{0}\Xi^{+}_5$.  
Notations are the same as in Fig.~\ref{nmset1}.}
\label{nmset2} 
\end{figure}   

In Fig.~\ref{nmset2}, we plot the total and differential cross-sections
for the negative-parity $\Xi_5$.  The energy dependence of the total
cross section looks similar to that for the positive-parity one.  
We find that $\sigma\sim 26\,nb$ for $F_{1}$ and $\sim 12\,nb$ for
$F_{2}$ in average for the CM energy region 
$E_{CM}^{\rm th}\,\le\,E_{CM}\,\le\,3.35\,{\rm GeV}$.  
We see that the total cross-sections are almost a hundred times
smaller than that for the positive-parity $\Xi_5$.  The difference
between the results of the two parities is even more pronounced than
in the previously investigated reactions, such as 
$\gamma N$, $K N$, and $N N$ 
scattering~\cite{Liu:2003rh,Hyodo:2003th,Oh:2003kw,nam1,nam2,nam3,
nam4,Liu:2003zi,Zhao:2003gs,Yu:2003eq,Kim:2003ay,
Huang:2003bu,Liu:2003ab,Li:2003cb},   
where typically the difference was about an order of ten.  
In the present reaction, the interference between the $s$- and the $u$-channels becomes important, in addition to the kinematical effect
in the p-wave coupling for the positive-parity (but not 
in the s-wave for the negative-parity), which is proportional to $\vec
\sigma \cdot \vec q$ and enhance the amplitude at high momentum
transfers.  In the case of the positive parity, the two terms which
are kinematically enhanced are interfered constructively, 
while for the negative-parity $\Xi_5$, the relatively 
small amplitudes without the enhancement factor is done destructively.   
These two effects are simultaneously responsible for the large
difference in the cross sections.  

In the right panel (b) of Fig.~\ref{nmset2}, the angular distributions
for the production of the negative-parity $\Xi_5$ are plotted.  Here,
the angular dependence changes significantly as compared with  
the positive-parity case.  When the form factor $F_2$ is used, 
forward scattering significantly increases because the bare amplitude 
shows an enhancement in the forward direction.  When using $F_1$,
however, due to its strong enhancement in the backward direction, 
the cross sections get quite larger in the backward direction, except
for those in the vicinity of the threshold, {\em i.e.}, $E_{CM} \le 2.45$ GeV.   

\subsection{$\bar{K}N\to K^{*}\Xi_5$}
In this subsection, we discuss the $K^*$ production. 
As explained in the previous section, the appearance of 
the  coupling constant $g_{K^* \Sigma \Xi_5}$ raises the problem of  
the relative sign in the amplitude.  First, we briefly discuss 
possible relations to determine the magnitude  
of the $g_{K^* \Sigma \Xi_5}$ coupling.  If we use the SU(3) relation 
this coupling may be set equal to $g_{K^* N \Theta}$.  There are
several discussions on the $g_{K^* N \Theta}$ 
coupling.  For example, a small value of the $g_{K^* N \Theta}$ was
chosen according to the relation $g_{K^{*}N\Theta}/g_{KN\Theta} = 1/2$  
as inferred from a phenomenological study of the hyperon coupling constants~\cite{Janssen:2001wk}, while in the
quark model, the decay of the pentaquark states  
predicts a positive parity $\Theta^+$ by using the relation  
$g_{K^{*}N\Theta}/g_{KN\Theta} =\sqrt{3}$~\cite{Close:2004tp}.  
In the meanwhile, we find $g_{K^{*}N\Theta}/g_{KN\Theta} =1/\sqrt{3}$
for the negative parity.  Since we are not able to determine the sign
of the coupling constant in this study, we will present the results
for four different cases: 
$g_{K^* \Sigma \Xi_5} = \pm \sqrt{3} g_{KN\Theta} = \pm 6.53$
and 
$g_{K^* \Sigma \Xi_5} = \pm 1/2 g_{KN\Theta} = \pm 1.89$ 
for positive parity,  and 
$g_{K^* \Sigma \Xi_5} = \pm \sqrt{3} g_{KN\Theta} = \pm 0.91$
and 
$g_{K^* \Sigma \Xi_5} = \pm 1/2 g_{KN\Theta} = \pm 0.27$ 
for negative parity.  

\begin{figure}[t]
\begin{tabular}{cc}
\includegraphics[width=7.5cm]{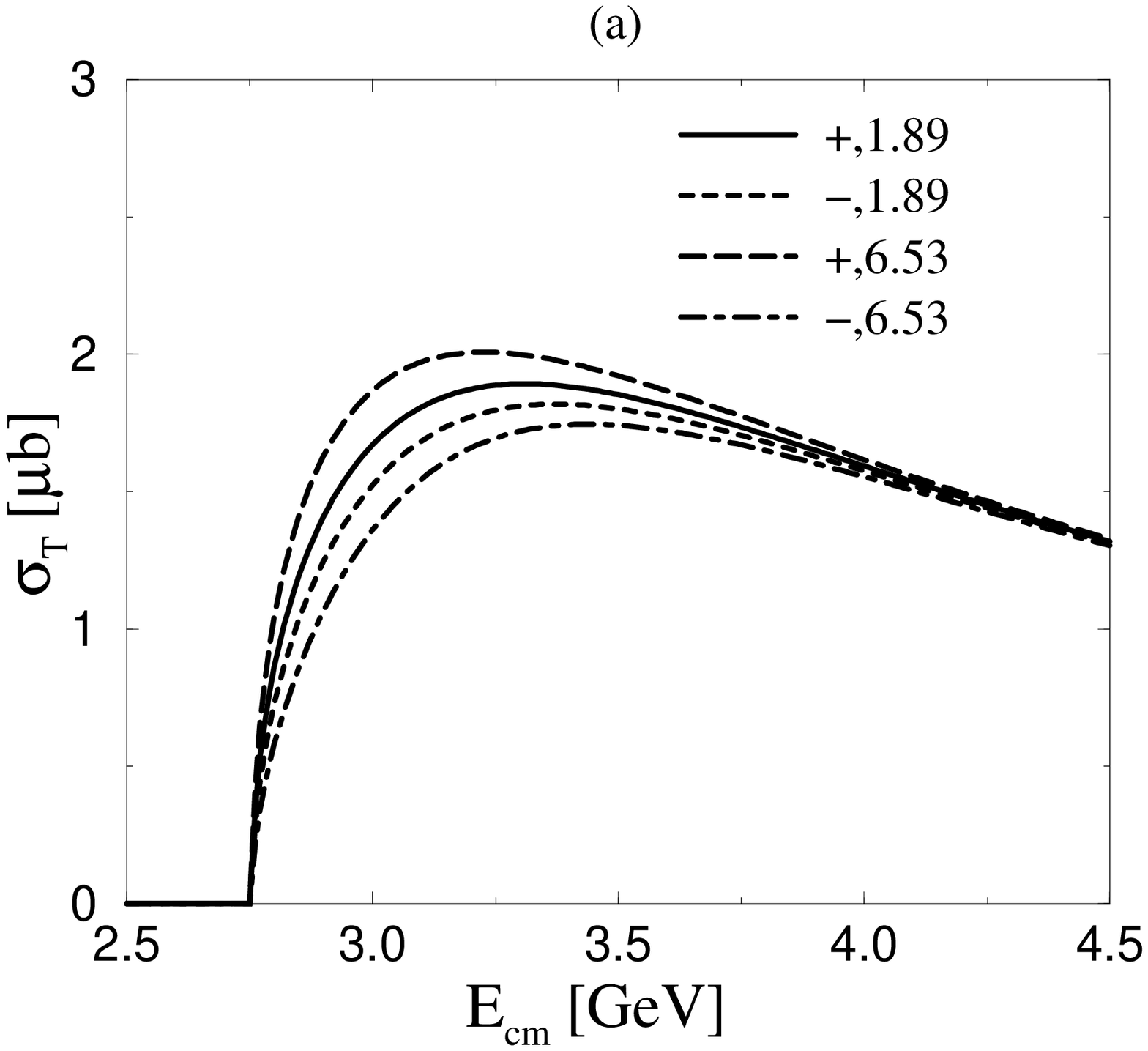}
\includegraphics[width=7.5cm]{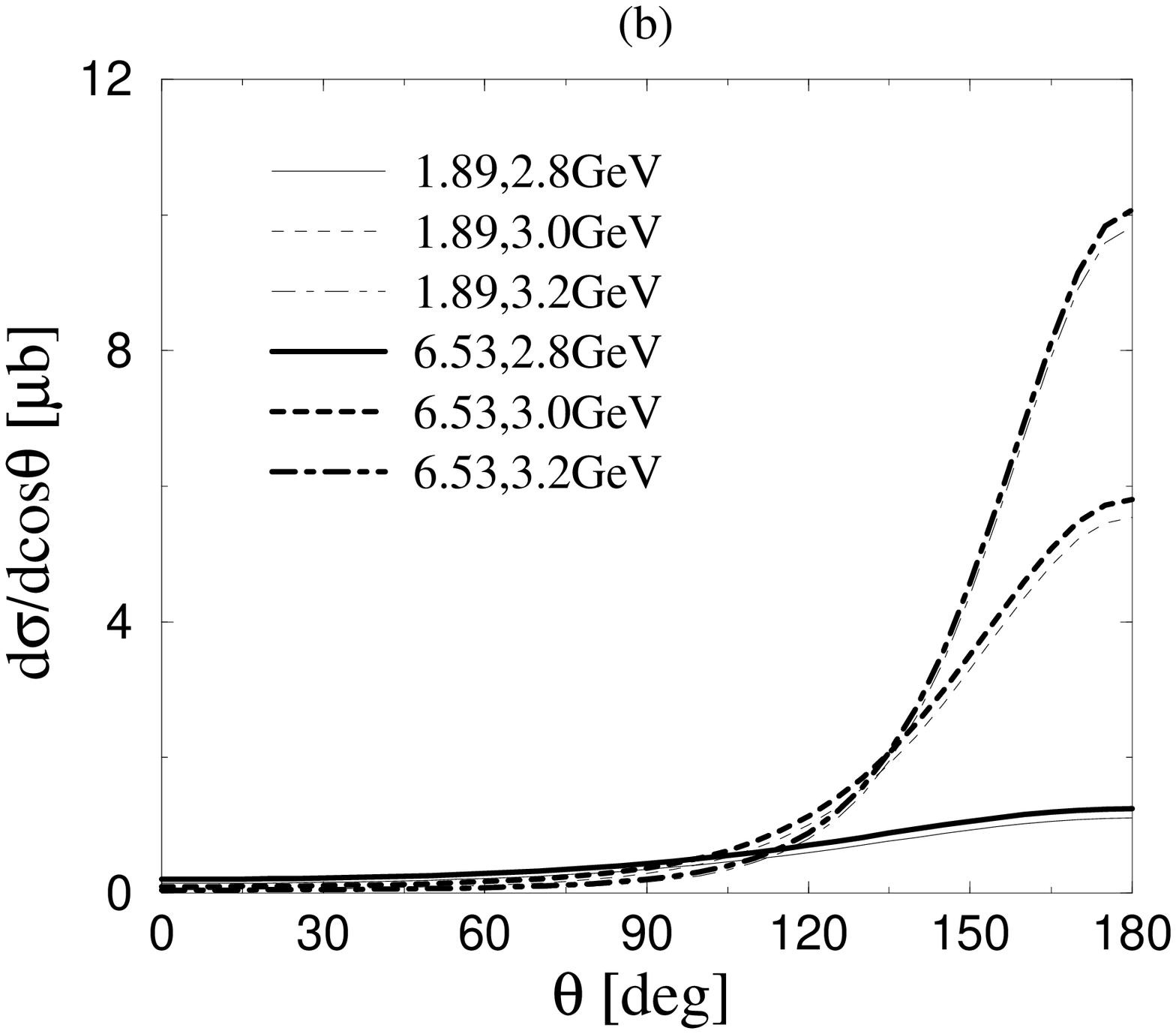}
\end{tabular}
\caption{Cross sections for production of the positive parity
$\Xi_5$ in the reaction $\bar{K}^{0}p\to K^{*0}\Xi^{+}_5$
with the $F_1$ form factor employed.  
The total cross-sections in (a) are calculated for four different 
$g_{K^* \Sigma \Xi_5}$ coupling constants as indicated by the 
labels. The angular distributions in (b) are calculated  for three different 
CM energies and two different $g_{K^* \Sigma \Xi_5}$ coupling 
constants as indicated by the labels.}
\label{nmset3}
\end{figure}

\begin{figure}[t]
\begin{tabular}{cc}
\includegraphics[width=7.5cm]{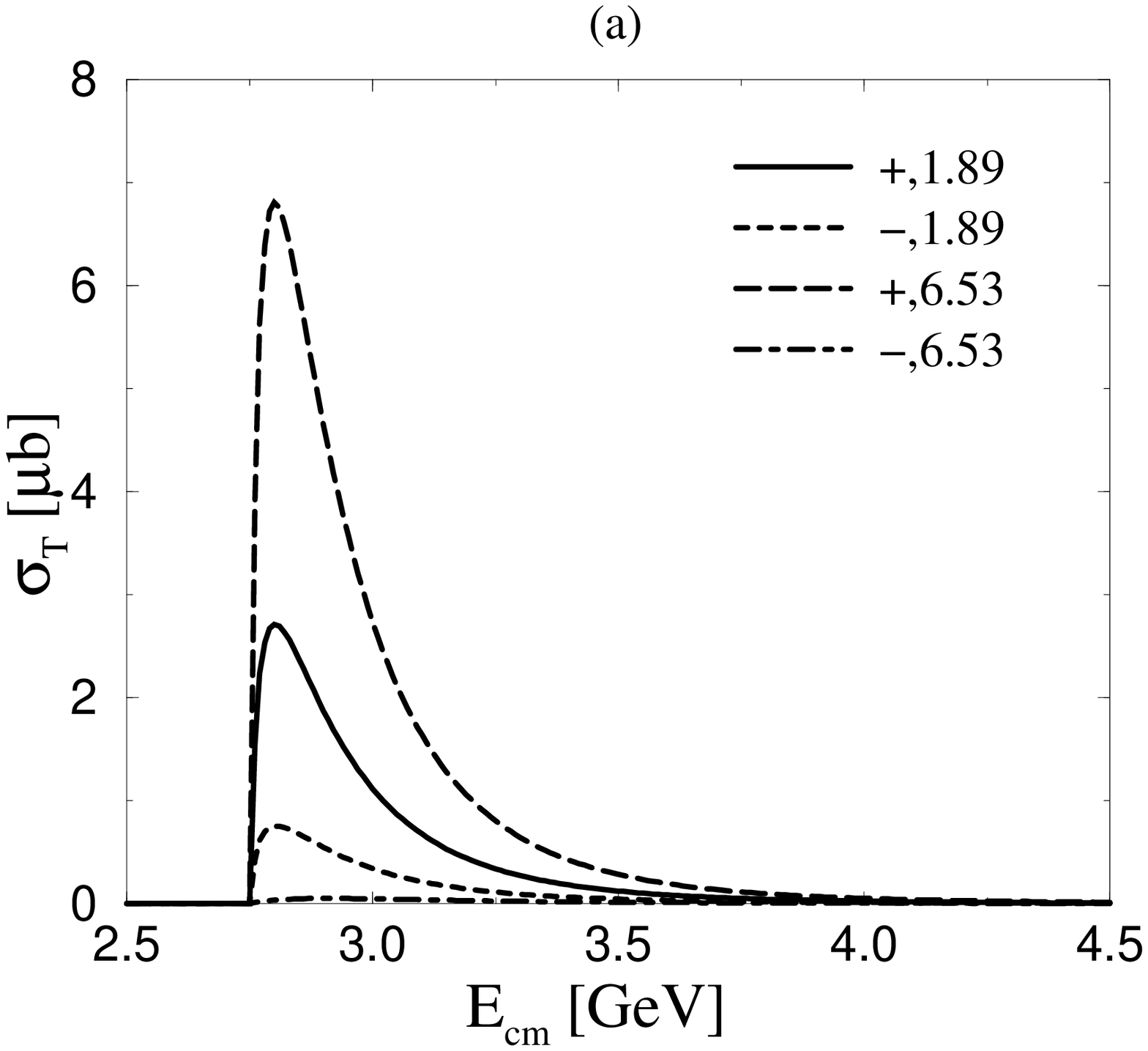}
\includegraphics[width=7.5cm]{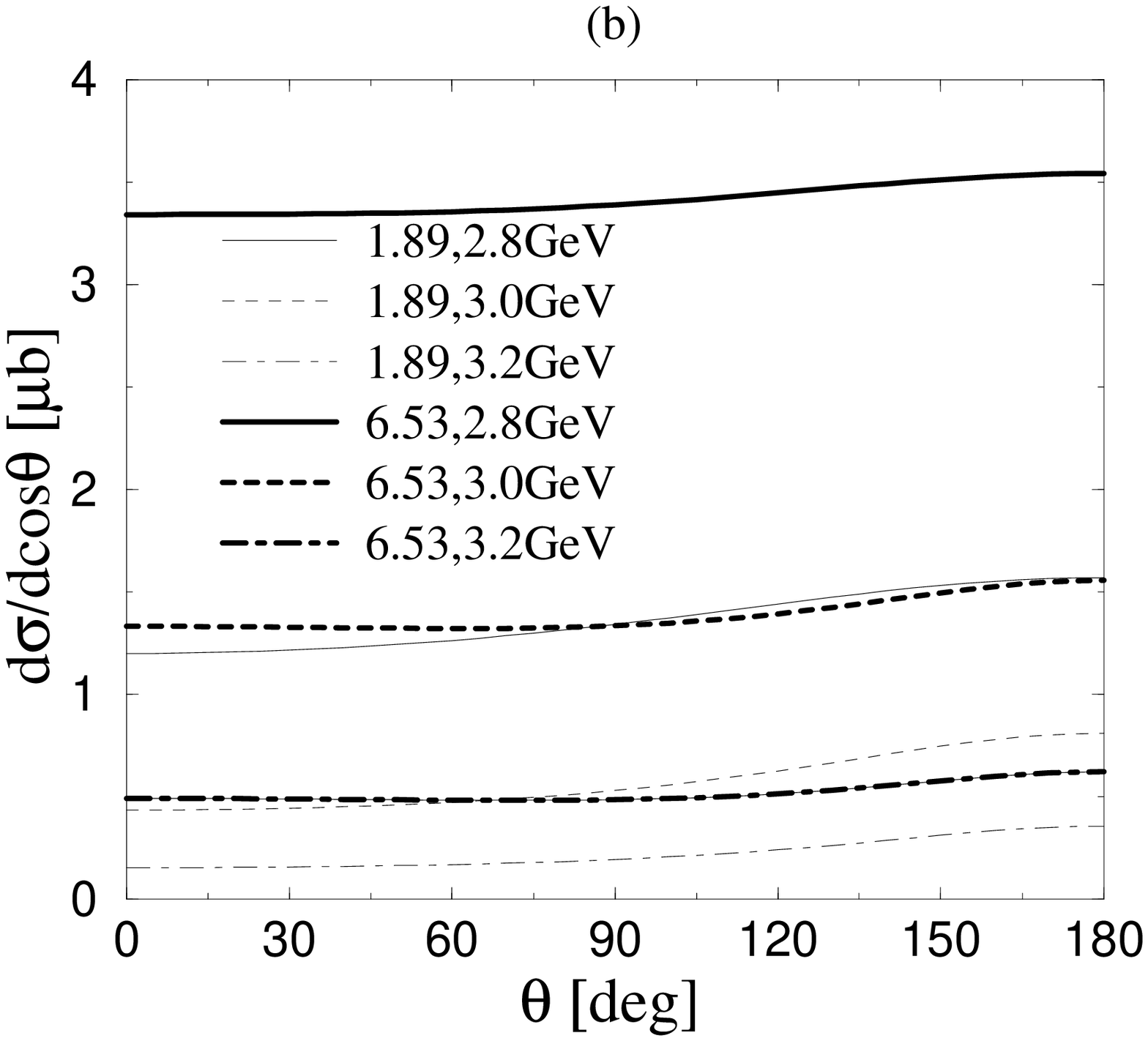}
\end{tabular}
\caption{Cross sections for production of the positive parity
$\Xi_5$ in the reaction $\bar{K}^{0}p\to K^{*0}\Xi^{+}_5$
with the $F_2$ form factor employed.  
For notations, see the caption of Fig.~\ref{nmset3}.}
\label{nmset4}
\end{figure}

Figures~\ref{nmset3} and \ref{nmset4} show the total and the differential
cross-sections for the positive parity $\Xi_5$ with the
$F_1$ and the $F_2$ form factors, respectively.  We present the results
with the four different coupling constants for the total cross-sections while for the differential cross-sections, we present those with
the two positive coupling constants at  
three different energies, $E_{CM} = 2.8$, $3.0$, and $3.2$ GeV.  The
results for the negative coupling constants are qualitatively  
similar to each other.  When the $F_1$ form factor is used, the results
do not depend on the choice of $g_{K^* \Sigma \Xi_5}$ because the
$u$-channel is the dominant component.  In this case, similar
discussions can be made as in the previous case of the 
$K$ production reaction.  
However, when $F_2$ is used, the results are very sensitive to the  
sign of $g_{K^* \Sigma \Xi_5}$, which determines whether
the $s$- and the $u$-channels interfere constructively or not. 

\begin{figure}[t]
\begin{tabular}{cc}
\includegraphics[width=7.5cm]{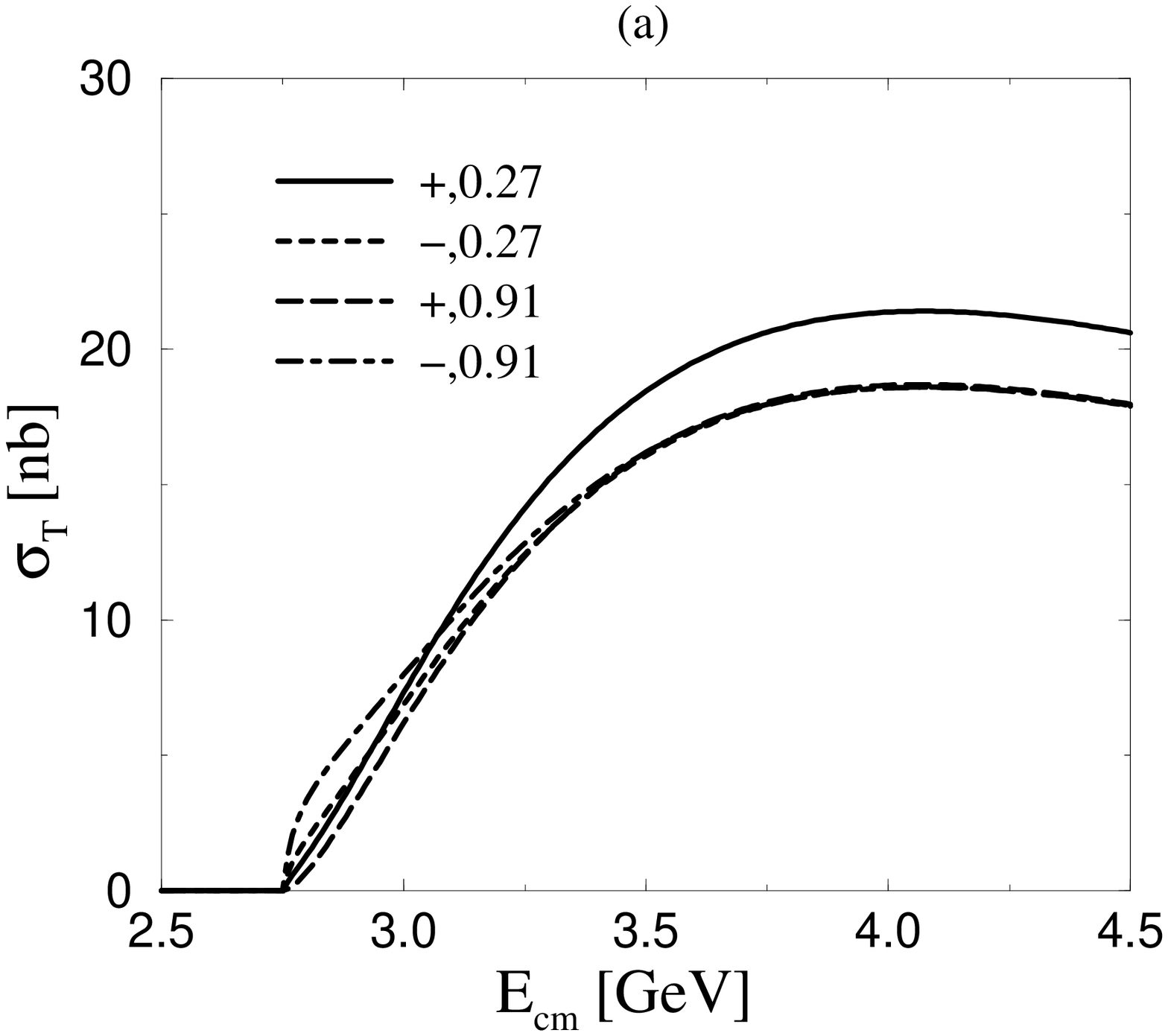}
\includegraphics[width=7.5cm]{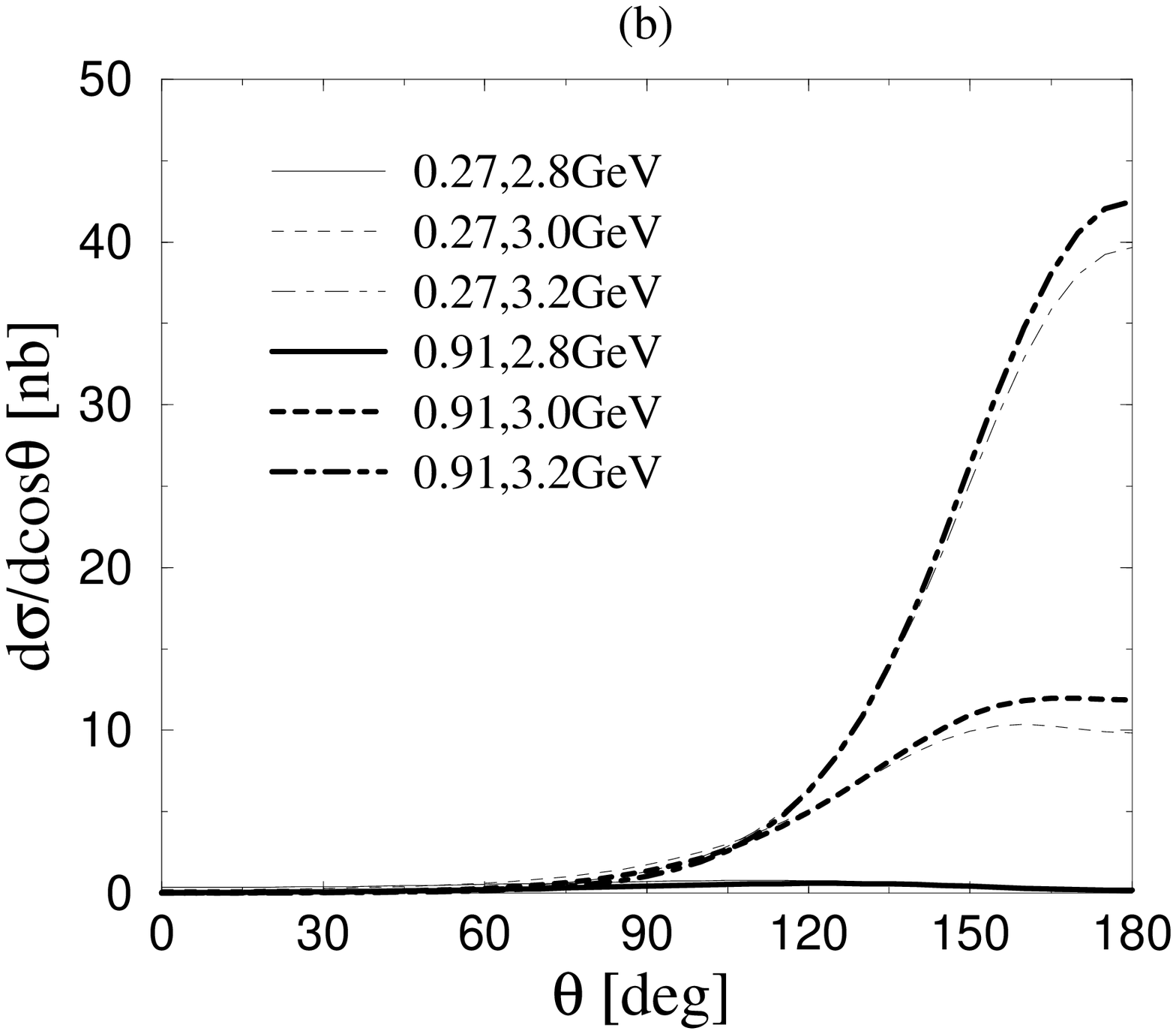}
\end{tabular}
\caption{
Cross sections for production of the negative parity
$\Xi_5$ in the reaction $\bar{K}^{0}p\to K^{*0}\Xi^{+}_5$
with the $F_1$ form factor employed.  
The total cross-sections in (a) are calculated for four different 
$g_{K^* \Sigma \Xi_5}$ coupling constants as indicated by the 
labels. The angular distributions in (b) are calculated  for three different 
CM energies and two different $g_{K^* \Sigma \Xi_5}$ coupling 
constants as indicated by the labels.}
\label{nmset5}
\end{figure}

\begin{figure}[tbh]
\begin{tabular}{cc}
\includegraphics[width=7.5cm]{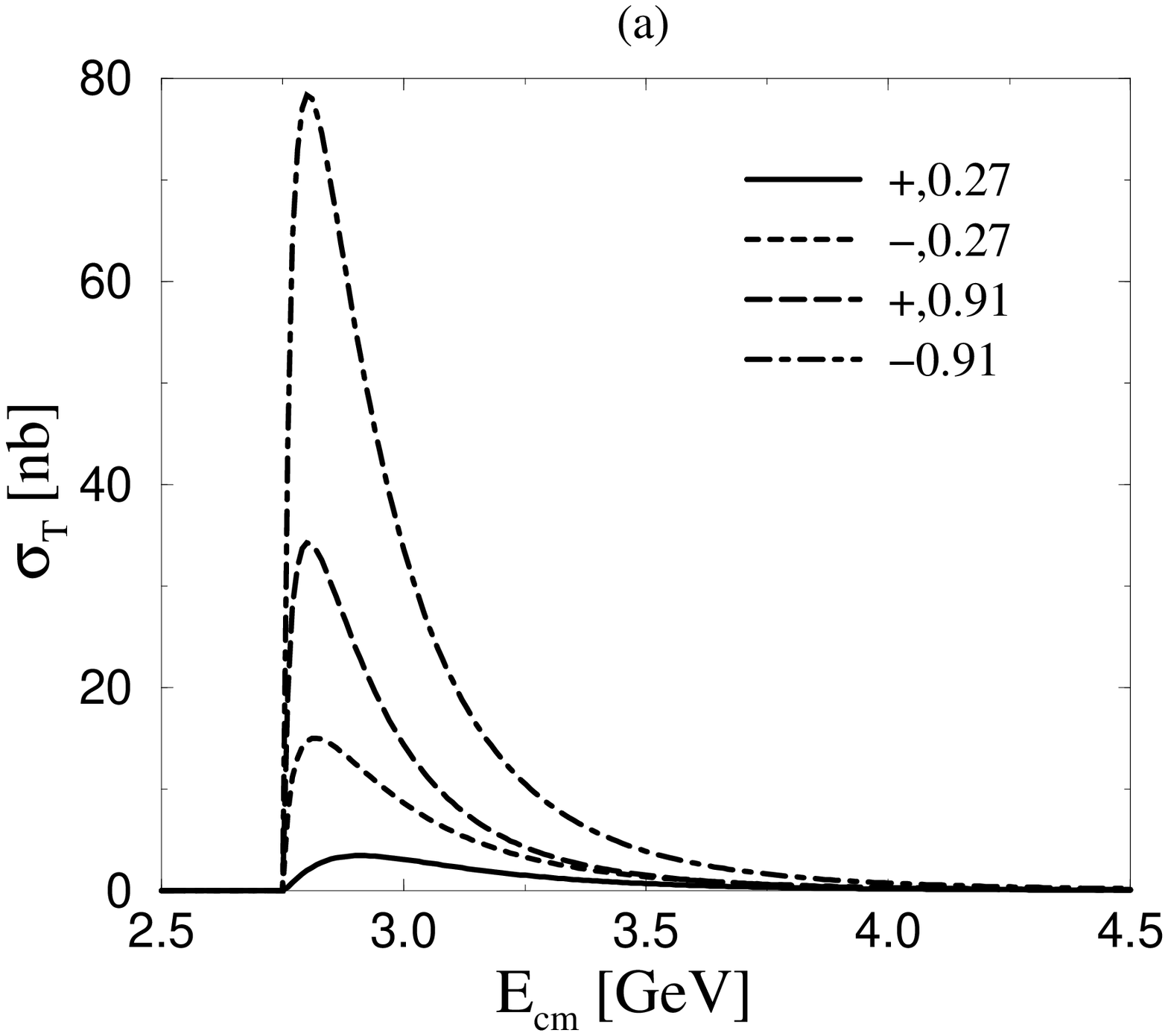}
\includegraphics[width=7.5cm]{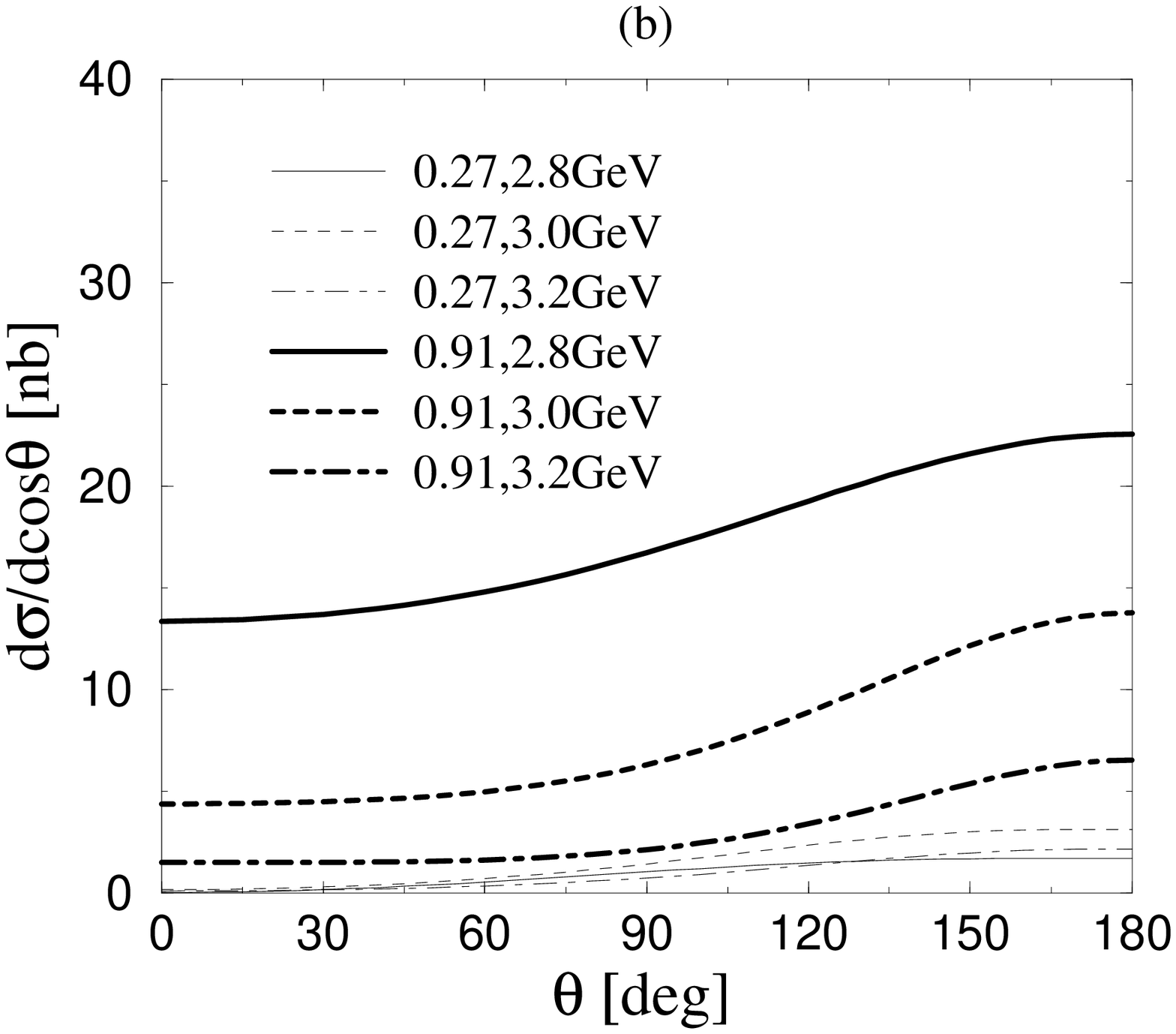}
\end{tabular}
\caption{
Cross sections for production of the negative parity
$\Xi_5$ in the reaction $\bar{K}^{0}p\to K^{*0}\Xi^{+}_5$
with the $F_2$ form factor employed.  
For notations, see the caption of Fig.~\ref{nmset5}.  
}
\label{nmset6}
\end{figure}

Fig.~\ref{nmset5} and Fig.~\ref{nmset6} show the results 
for the negative-parity $\Xi_5$ with the $F_1$ and the $F_2$ form factors used, respectively. 
Similar discussions apply for this case 
as for the positive parity case, but the values of cross sections
are reduced by about a factor of a hundred.  
\section{Summary and Discussion}
We have studied the production of the pentaquark exotic baryon
$\Xi_5\,({\rm mass} = 1862\,{\rm MeV},\,I = 3/2\,S = -2,\,{\rm spin} =
1/2\,({\rm assumed}))$ in the  reactions $\bar{K}N\to K\Xi_5$ and 
$\bar{K}N\to K^{*}\Xi_5$.  We have employed two different
phenomenological form factor Eqs.~(\ref{ff1}) and (\ref{ff2}),
with appropriate parameters for the coupling strengths and the cutoff parameters. 
In the present reactions, since two units of strangeness 
are transferred, 
only $s$- and $u$- channel diagrams are allowed at the tree level.  
On one hand, this fact simplifies the reaction mechanism 
and, hence, the computation.  
Furthermore, there is no ambiguity in the relative signs of 
coupling constants for the case of $K$ production.  
On the other hand, the cross sections strongly depend on the 
choice of form factors.  
In fact, Fig.~\ref{nmset1}$\sim$\ref{nmset6} show that we have found a rather different energy and angular 
dependence when using different form factors.  
At this moment, it is difficult theoretically to 
say which is better.  
Nevertheless, it would be useful to summarize the present 
result for the total cross-sections in Table~\ref{table1}. 
There, we see once again that the total cross-sections are,
generally, much larger for positive-parity $\Xi_5$ 
than for positive-parity one by about factor of a hundred because there is a
cancellation due to destructive interference. 
\begin{table}[b]
\begin{tabular}{c|cc||c|cc}
Reaction & $F_{1}$ & $F_{2}$ & Reaction & $F_{1}$ & $F_{2}$\\
\hline
$\sigma_{\bar{K}N\to K\Xi_5}(P = +1)$ & 2.6 $\mu b$&
  1.5 $\mu b$ & $\sigma_{\bar{K}N\to K^{*}\Xi_5}(P =
    +1)$ & 1.6 $\mu b$ & $\lesssim$ 2  $\mu b$ \\
$\sigma_{\bar{K}N\to K\Xi_5}(P = -1)$ & 26 $nb$ & 12
  $nb$ &$\sigma_{\bar{K}N\to K^{*}\Xi_5}(P = -1)$ &
    14  $nb$& $\lesssim$ 20 $nb$
\end{tabular}
\caption{Summary for 
the average total cross-sections in the CM energy region
$2.35\,{\rm GeV}\,\le\,E_{\rm CM}\,\le\,3.35\,{\rm GeV}$ for
$\bar{K}N\to K\Xi_5$ and  $2.75\,{\rm
GeV}\,\le\,E_{\rm CM}\,\le\,3.75\,{\rm GeV}$ for
$\bar{K}N\to K^{*}\Xi_5$.  
For $K^*$ production with the $F_2$ form factor used, only the 
upper values are quoted because the interference  
suppresses them.}
\label{table1}
\end{table}   
\section*{Acknowledgements}
The authors are supported by a Korea Research
Foundation grant (KRF-2003-041-C20067).  The work of S.i.N. was supported by a scholarship endowed by the Ministry of Education,
Science, Sports and Culture of Japan. He is also supported by a grant
for Scientific Research (Priority Area No. 17070002) from the Ministry
of Education, Culture, Science and Technology, Japan.

\end{document}